\documentclass{CUP-JNL-DTM}

\usepackage{graphicx}
\usepackage{subcaption}
\usepackage{caption}
\usepackage{multicol,multirow}
\usepackage[figuresright]{rotating}
\usepackage{array}
\usepackage{geometry}

\usepackage{booktabs}
\usepackage{tabularx}

\usepackage{appendix}
\usepackage{ifpdf}
\usepackage[T1]{fontenc}
\usepackage[utf8]{inputenc}
\usepackage[backend=biber,style=numeric-comp,sorting=none,maxnames=3,minnames=2]{biblatex}
\usepackage{textcomp}
\usepackage{xcolor}
\definecolor{latent-purple}{RGB}{89,85,215}
\usepackage{nicematrix}
\usepackage{seqsplit}
\usepackage{tikz}
\usepackage[section]{placeins}

\usepackage{amsthm}
\usepackage{amsmath,amssymb,amsfonts}
\usepackage{newtxtext}
\usepackage{newtxmath}

\theoremstyle{definition}

\numberwithin{equation}{section}
\usepackage{diagbox}
\usepackage{float}

\usepackage[colorlinks,allcolors=latent-purple]{hyperref}
\usepackage{cleveref}
\crefname{appendix}{App.}{Apps.}
\crefname{section}{Sec.}{Secs.}
\crefname{subsection}{Sec.}{Secs.}
\crefname{subsubsection}{Sec.}{Secs.}
\crefname{figure}{Fig.}{Figs.}
\crefname{table}{Tab.}{Tabs.}
\crefname{equation}{Eq.}{Eqs.}
\crefname{algorithm}{Alg.}{Algs.}
\crefname{example}{Ex.}{Exs.}
\crefname{page}{p.}{pp.}
\crefname{line}{l.}{ll.}

\usepackage{siunitx}
\DeclareSIUnit\angstrom{\text{Å}}
\DeclareSIUnit\molar{\textsc{M}}
\DeclareSIUnit\rpm{rpm}

\usepackage[acronym]{glossaries}
\makeglossaries
\glsdisablehyper

\newacronym{spr}{SPR}{surface plasmon resonance}
\newacronym{htspr}{HT-SPR}{high-throughput surface plasmon resonance}
\newacronym[
    plural={T-cell engagers},
    firstplural={T-cell engagers (TCEs)}
]{TCE}{TCE}{T-cell engager}
\newacronym{igg}{IgG}{immunoglobulin G}
\newacronym{scfv}{scFv}{single-chain variable fragment}
\newacronym{vhh}{VHH}{variable heavy-chain-only}
\newacronym{fv}{Fv}{fragment variable}
\newacronym{cdr}{CDR}{Complementarity-Determining Region}
\newacronym{pdb}{PDB}{Protein Data Bank}
\newacronym{sabdab}{SAbDab}{The Structural Antibody Database}
\newacronym{bli}{BLI}{bio-layer interferometry}
\newacronym{pbmc}{PBMC}{peripheral blood mononuclear cells}
\newacronym{phd2}{PHD2}{prolyl hydroxylase domain-containing protein 2}

\newacronym{il6}{IL-6}{interleukin-6}
\newacronym{il6r}{IL-6R}{interleukin-6 receptor}
\newacronym{il33}{IL-33}{interleukin-33}
\newacronym{prl}{PRL}{prolactin}
\newacronym{prlr}{PRLR}{prolactin receptor}
\newacronym{tnfl9}{TNFL9}{TNF ligand superfamily member 9}
\newacronym{tnfl9r}{TNFRSF9}{TNF receptor superfamily member 9}
\newacronym{tnfa}{TNF$\alpha$}{tumour necrosis factor alpha}
\newacronym{tnfr1}{TNFR1}{TNF receptor 1}
\newacronym{htfr}{hTfR1}{human transferrin receptor 1}
\newacronym{sc2rbd}{SC2RBD}{SARS-CoV-2 receptor-binding domain}
\newacronym{ace2}{ACE2}{angiotensin-converting enzyme 2}
\glsunset{il6}\glsunset{il6r}\glsunset{il33}
\glsunset{prl}\glsunset{prlr}
\glsunset{tnfl9}\glsunset{tnfl9r}\glsunset{tnfa}\glsunset{tnfr1}
\glsunset{htfr}\glsunset{sc2rbd}\glsunset{ace2}

\usepackage{fancyhdr}
\fancyheadoffset{0pt}
\newcommand{\papertitle}{\raggedright Latent-Y: A Lab-Validated
Autonomous Agent for De Novo Drug Design}

\usepackage[version=4]{mhchem}

\newcommand{\themodel}{\mbox{Latent-Y}}

\newcommand{\lxtwo}{{Latent-X2}}
\newcommand{\llplatform}{{Latent Labs Platform}}

\newcommand{\denovo}{\textit{de novo}}

\newcommand{\insilico}{\textit{in silico}}

\newcommand{\kd}{\ensuremath{\mathrm{K}_{\mathrm{D}}}}
\newcommand{\kon}{\ensuremath{\mathrm{k}_{\mathrm{on}}}}
\newcommand{\koff}{\ensuremath{\mathrm{k}_{\mathrm{off}}}}
\newcommand{\rmax}{\ensuremath{\mathrm{R}_{\mathrm{max}}}}

\newcommand{\subfigref}[2]{Fig.~\hyperref[#1]{\ref*{#1}#2}}

\newcommand{\customcaption}[1]{\refstepcounter{figure}\caption*{\textbf{Fig.~\thefigure{}~|}~#1}}

\newcommand{\customtablecaption}[1]{\refstepcounter{table}\caption*{\textbf{Tab.~\thetable{}~|}~#1}}

\newcommand{\ArtType}{}
\makeatletter
\newcommand{\@opjournalheader}{}
\makeatother
\newcommand{\headeright}{}

\DeclareUnicodeCharacter{03B1}{\ensuremath{\alpha}}
\DeclareUnicodeCharacter{03B2}{\ensuremath{\beta}}

\usepackage{tcolorbox}

\begin{document}

\setcounter{figure}{0}
\setcounter{table}{0}

\begin{Frontmatter}
    \title[Article Title]{\papertitle
    }

    \author[]{Latent Labs Team}
    \address[]{\orgaddress{\city{London, UK \& San Francisco, USA}\\
        \orgaddress{23 March 2026}} \\
        \begin{center}
            \makebox[\textwidth][c]{\includegraphics[width=1.0\textwidth]{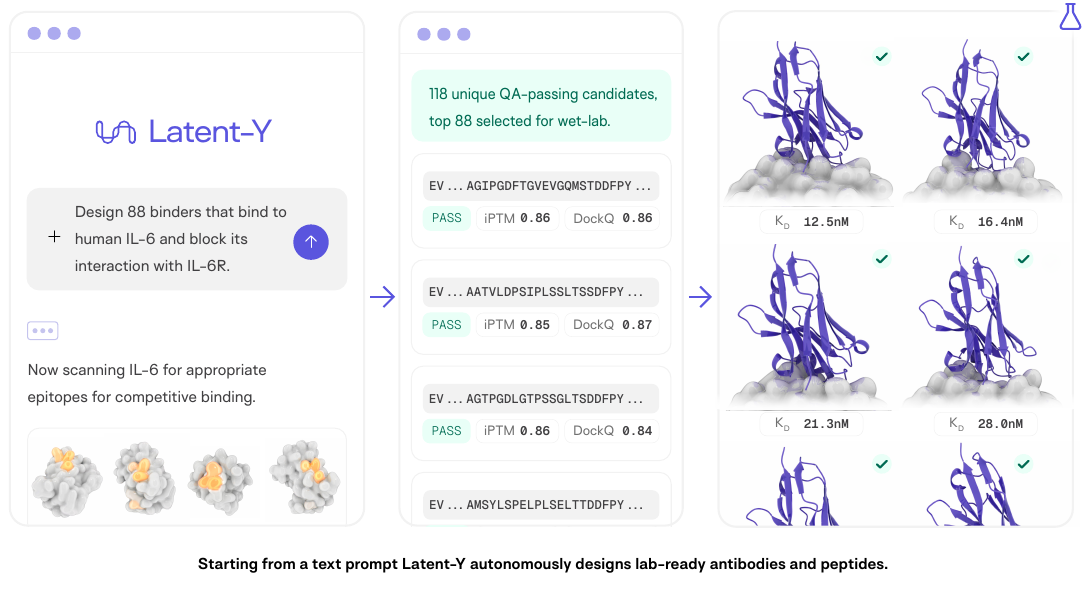}}
        \end{center}
        \vspace{-1cm}
    }

    \abstract{
        Drug discovery relies on iterative expert workflows that are
        slow to parallelize and difficult to scale. Here we introduce
        \themodel{}, an AI agent that autonomously executes complete
        antibody design campaigns from text prompts, covering
        literature review, target
        analysis, epitope identification, candidate design,
        computational validation, and selection of lab-ready
        sequences. \themodel{} is
        integrated into the
        \llplatform{}, where it operates in the same environment as
        drug-discovery experts with access to bioinformatics tools,
        biological databases, and scientific literature. The agent can
        run fully autonomously end-to-end, or collaboratively, where
        researchers review progress, provide feedback, and direct
        subsequent steps. Candidate antibodies are generated using
        \lxtwo{}, our frontier generative model for drug-like
        antibody design. We demonstrate the agent's capability across
        three distinct campaign types: epitope discovery
        guided by therapeutic specifications,
        cross-species binder design, and autonomous design from a
        scientific publication targeting human transferrin receptor for
        blood--brain barrier crossing. Across nine targets,
        \themodel{} produced lab-confirmed nanobody binders against
        six, achieving a \SI{67}{\percent} target-level success rate
        with binding affinities reaching the single-digit nanomolar range,
        without human filtering or intervention. In user studies,
        experts working with \themodel{} completed design campaigns
        56 times faster than independent expert time estimates,
        compressing weeks of work into hours. Because \lxtwo{} is a
        general-purpose atomic-level model for biologics design, the
        same agent architecture naturally extends to macrocyclic
        peptide and mini-binder design campaigns, broadening
        autonomous discovery across therapeutic modalities.
        \themodel{} is available to selected partners at
        \href{https://platform.latentlabs.com}{\texttt{\color{latent-purple}platform.latentlabs.com}}.
    }
\end{Frontmatter}

\section[Introduction]{Introduction}

Frontier AI models such as \lxtwo{}~\autocite{lx2} and other molecular design
models~\autocite{chai2025drug, nabla2025novo} have recently demonstrated the
feasibility of zero-shot biologics design, directly producing antibody
and peptide candidates that have drug-like properties and advancing prior
work in protein structure prediction~\autocite{jumper2021highly,
abramson2024accurate} and generative
protein design~\autocite{dauparas2022robust, watson2023novo,
    zambaldi2024novo, pacesa2024bindcraft, lx1, hayes2025simulating,
geffner2025proteina, stark2025boltzgen, team2025pxdesign}.

In a world in which we can now design lab-ready molecules on the
computer, the rate-limiting step for early drug discovery is
no longer finding a candidate molecule for a drug program. The rate-limiting
step instead becomes the bandwidth of drug-discovery
organizations and access to the PhD-level domain expertise required to
do so at scale.

To address this bottleneck, we developed \themodel{}, an agentic
system for \denovo{} drug
design, capable of designing antibodies, peptides, and mini-binders
from text descriptions of drug design goals. \themodel{} directly
addresses the exploration and scale bottleneck by delivering
autonomous drug design research that can be executed in parallel. We
demonstrate, for the first time, that an autonomous agent can design
lab-confirmed \denovo{} antibody binders purely from text descriptions
of goals and requirements, without requiring manual intervention in
the design process. \mbox{\themodel{}-designed} nanobodies display strong
binding affinities reaching single-digit nanomolar ranges, with highly
competitive target-level success rates, hitting 6 of 9 targets
attempted in the lab.

Where \lxtwo{} reasons at the atomic level to design precise molecular
interactions, \themodel{} operates in the same environment as human
experts, applying expert-level reasoning to navigate from research
objective to lab-ready candidates. It queries molecular and literature
databases, orchestrates \lxtwo{} to generate and score candidates, and
employs bioinformatics tools to analyse
the sequence and structure of targets, epitopes, and designed
molecules. It intelligently explores a large combinatorial design space. It
tests different hypotheses, analyses results, and integrates
insights across parallel runs, plans appropriate next steps, and works
towards user-provided goals, such as achieving an intended functional
mechanism or producing the requested number of lab-ready designs.

In this work, \themodel{} is run both fully autonomously and
collaboratively. The majority of campaigns were executed end to end
without human intervention, while the cross-species campaign
demonstrates tight human-agent collaboration, with the researcher
steering strategy and the agent adapting in real time. Both modes are
supported across the full spectrum of expert involvement. We show that
\themodel{} significantly accelerates the work of protein design experts,
reducing the time required for expert research and computational
workflows from weeks to hours. This acceleration compounds when
running \themodel{} instances at scale and in parallel, for example to
find development candidates across multiple pre-clinical drug programs
simultaneously.

\textbf{Our main contributions are:}
\begin{enumerate}
    \item The first autonomous agent for \denovo{} biologics design,
        delivering lab-ready sequences from text input.
    \item Lab-validated \denovo{} nanobody design from text, with a
        \SI{67}{\percent} target-level success rate across 9 targets.
    \item \themodel{}-designed \denovo{} nanobodies with single-digit
        nanomolar binding affinities, confirmed in the lab.
    \item A 56-fold acceleration of expert-led design campaigns
        with \themodel{} versus without, measured against independent
        expert time estimates.
    \item Lab-validated cross-species binder design via
        \themodel{}-generated custom generative code.
\end{enumerate}

\noindent\textbf{Availability.} \themodel{} is integrated and
available on the \llplatform{} at
\href{https://platform.latentlabs.com}{\texttt{\color{latent-purple}platform.latentlabs.com}}.

\begin{figure}[H]
    \centering
    \includegraphics[width=0.99\textwidth]{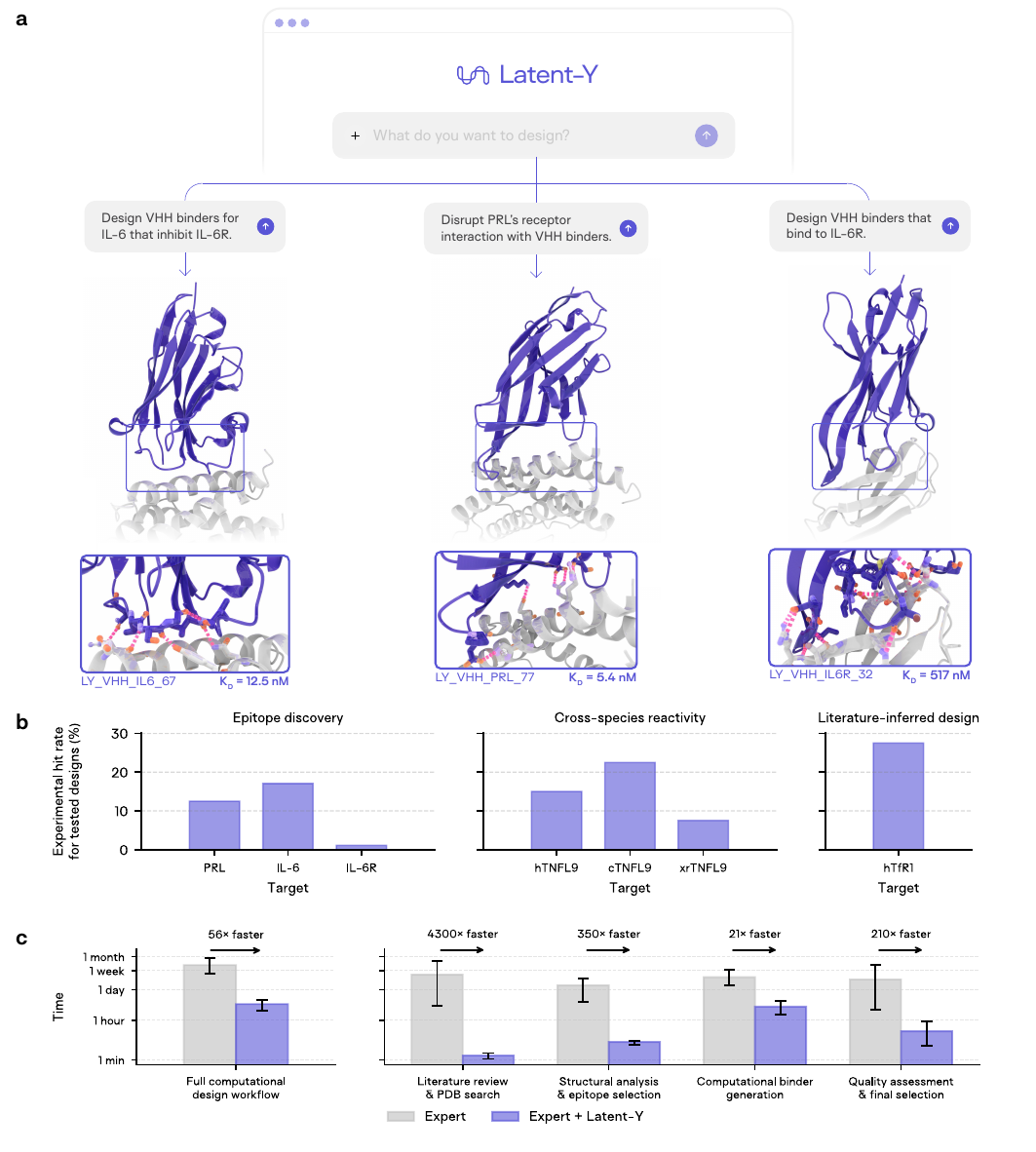}
    \customcaption{\textbf{\themodel{} autonomously designs
            nanomolar-affinity antibodies from text prompts, accelerating
        expert workflows by over 50-fold.}
        \textbf{(a)} Three independent VHH design campaigns, each
        initiated from a single natural-language prompt targeting
        \gls{il6}, \gls{prl}, and \gls{il6r} respectively. Designed
        structures of the top binder--target complex are shown alongside
        binding interface detail, with insets highlighting designed
        non-covalent interactions at the binding interface (pink dashed
        lines). All designs were experimentally characterized by
        \gls{spr}.
        \textbf{(b)} Per-target experimental hit rates across all
        successful targets. cTNFL9, cynomolgus \gls{tnfl9}; hTNFL9, human
        \gls{tnfl9}; xrTNFL9, cross-reactive \gls{tnfl9}.
        \textbf{(c)} Estimated time required to complete a full
        computational protein design campaign, comparing expert alone
        versus expert assisted by \themodel{}. Expert baseline times were
        obtained by polling independent protein designers across academia
        and industry ($n = 10$); \themodel{}-assisted times reflect the
        average across agent-assisted runs ($n = 5$). Error bars indicate
        minimum and maximum values.
        \textbf{(d)} Time breakdown by major design stage, including
        literature review and \gls{pdb} search, structural analysis and
        epitope selection, computational binder generation, and quality
        assurance and final selection. Error bars indicate minimum and
    maximum values.}
    \label{fig:summary}
\end{figure}

\section[Results]{Results}
\label{sec:results}

We evaluate \themodel{} across three antibody design settings:
epitope discovery campaigns targeting
\gls{il6}~\autocite{Tanaka_IL6therapeutic_2014},
\gls{prl}~\autocite{bernard_ProlactinReview_2019},
\gls{il33}~\autocite{liew_IL33_Health_Disease_2016},
\gls{tnfa}~\autocite{brenner_TNFSignaling_2015}, and
\gls{sc2rbd}~\autocite{lan_SC2RBDAce2Structure_2020}, with the goal
of therapeutic inhibition, and
\gls{il6r}~\autocite{barille_IL6RMultipleMyeloma_2000}, with the goal
of target binding; cross-species
reactivity campaigns against \gls{tnfl9}~\autocite{wang2009immune};
and literature-inferred
design campaigns targeting
\gls{htfr}~\autocite{gammella_TfR1IronGate_2017}. Across these campaigns,
\themodel{} achieves a target-level success rate of
\SI{67}{\percent}, with per-target hit rates ranging from
\SIrange{1}{28}{\percent} of tested sequences, and binding affinities
reaching the single-digit nanomolar range.

\subsection{\themodel{} is a wet-lab-validated agent for drug design
at scale}
\label{sec:agent_overview}

\themodel{} is natively integrated into the \llplatform{} and
can be accessed entirely through a browser, steered by natural language. A
researcher can provide a therapeutic objective as free text, a
structured work plan specifying targets and constraints, or a
scientific publication from which the agent autonomously extracts
target identities, biological context, and
design constraints, as shown in \subfigref{fig:summary}{a}.

Drug design problems are inherently underspecified at the outset: the
right epitope, the appropriate structural context, and the relevant
constraints often only become clear during the research process
itself. \themodel{} is designed to reduce this ambiguity. Upon receiving a
design objective, it consults scientific literature and databases to
build biological context, identifies appropriate target structures,
and characterizes candidate epitopes against functional criteria.
Unlike fixed generative pipelines, it accepts and reasons about
arbitrary research constraints such as modality, intended function,
species requirements, structural considerations, and computational
metrics and predictions, translating high-level goals into concrete
design decisions. It spawns targeted computational experiments using
\lxtwo{}, reasons about results, and doubles down on the most
productive directions, adjusting generation parameters, modifying
epitope constraints, or re-routing to alternative modalities as
needed. Where standard capabilities are insufficient, the agent can
generate custom computational approaches from natural language
descriptions, extending its own toolkit to address novel design
challenges, as described in \cref{sec:crossspecies}. Outputs
annotated with the agent’s reasoning are
produced at each stage for researcher inspection and, where needed,
manual override. Before
finalizing candidates, the agent performs quality assurance on the
designs, including clustering for diversity, sequence similarity
searches against external databases, and sequence liability analyses,
selecting a final set of lab-ready designs that satisfy the
user-defined objectives. Representative condensed reasoning traces illustrating
fully autonomous and collaborative human--agent workflows are shown in
\cref{fig:il6_trace} and \cref{fig:tnfl9_trace} respectively.

At the level of an individual campaign, \themodel{} compresses the
full computational antibody design workflow approximately 56-fold, on
average reducing
two weeks of expert effort to five hours, as shown in
\subfigref{fig:summary}{c}. When a researcher runs multiple
campaigns in parallel, these gains compound further still, multiplying
the effective throughput of a single researcher beyond what any
individual could achieve sequentially. The largest accelerations arise
in the reasoning-intensive stages: literature review and \gls{pdb}
analysis ($\sim$4,300$\times$) and structural analysis and epitope
selection ($\sim$350$\times$), as detailed in
\subfigref{fig:summary}{d}. Baseline
expert times were estimated by polling independent, PhD-level protein designers
across academia and industry, with a median of
9.8 years of relevant experience, as described in
\cref{app:methods_survey}. These
gains compound further at scale, with a single researcher able to
complete in under a day computational work that would previously have
taken several months, shifting drug
design from a sequential expert workflow to a scalable parallel
process limited by scientific ideas rather than time.

\begin{figure}[H]
    \centering
    \includegraphics[width=0.99\textwidth]{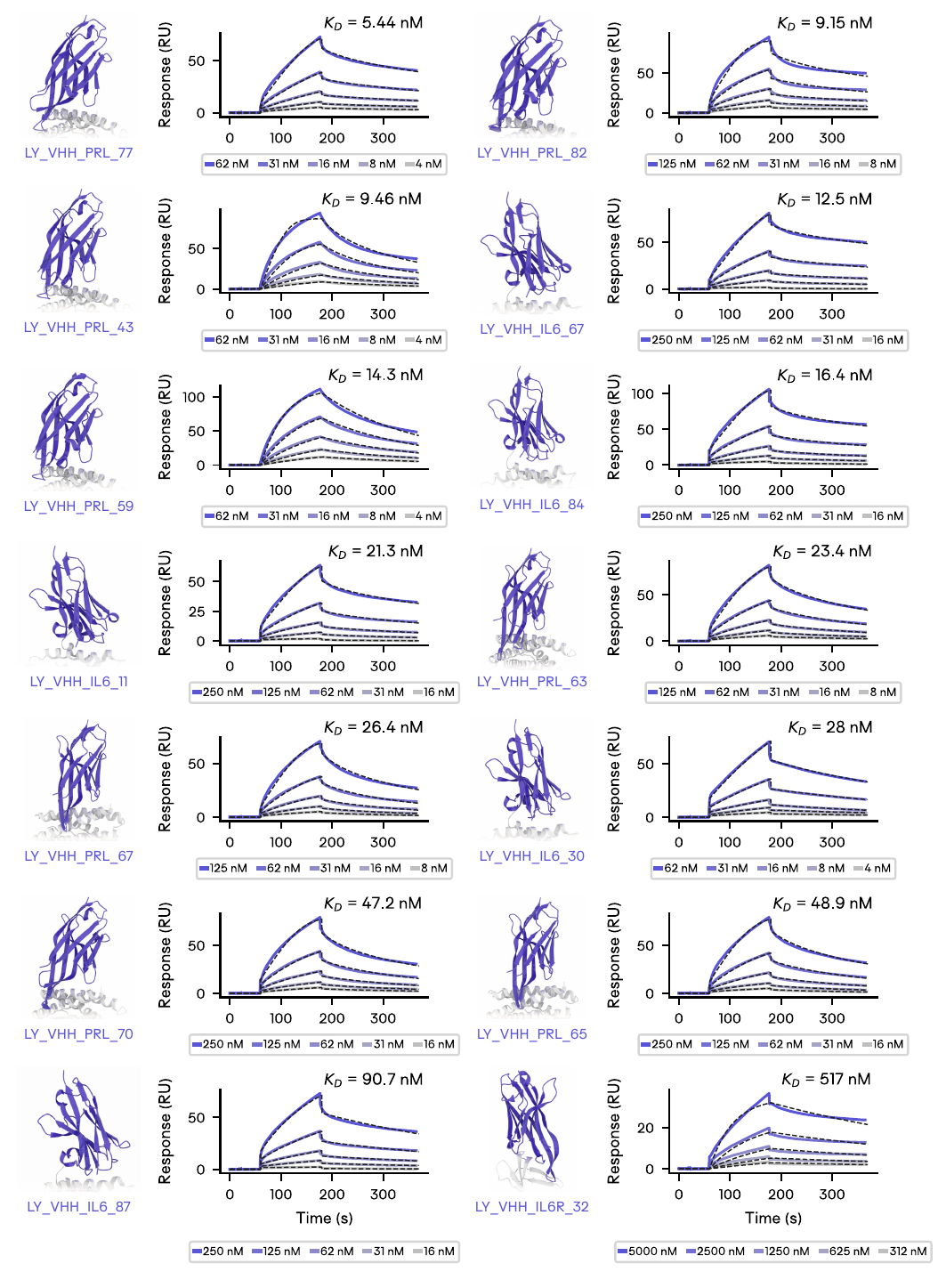}
    \customcaption{\textbf{Biophysical characterization of the best
            \themodel{}-designed VHHs against \gls{il6}, \gls{il6r}, and
            \gls{prl}, with affinities reaching the single-digit nanomolar
        range.} Top-performing \denovo{} VHH binders ranked by binding
        affinity, with corresponding designed bound structures and
        \gls{spr} response curves. Binding affinities were measured
        using five analyte concentrations with kinetic fitting to
        determine \kd{}, see~\cref{sec:methods_spr}. Reported \kd{}
        values span the nanomolar
        range, with lower \kd{} values corresponding to stronger
    binding.}
    \label{fig:sensograms}
\end{figure}

\subsection{\themodel{} enables autonomous design of low nM-affinity
antibodies}
\label{sec:wetlab_results}

In the epitope discovery setting, we evaluate \themodel{} by running
fully autonomous \denovo{} VHH design campaigns against six
therapeutically relevant targets, experimentally validating the
resulting sequences in the wet lab, see \cref{fig:sensograms,tab:reagents}.
Campaigns were specified via natural language prompts, with
prompts specifying only the high-level objective and
leaving all design decisions to the agent. For the inhibition-focused
campaigns, \themodel{} selected epitopes consistent with the intended
mechanism of action, indicating that the agent reliably identifies
functionally relevant binding sites from high-level goals alone.
For \gls{il6r}, given an unconstrained binding objective, the agent
identified an epitope on domain~1 of the receptor ectodomain, a region
absent from most co-crystal structures.

Successful campaigns produced experimentally confirmed VHH binders
against three therapeutically relevant targets:
\gls{il6}~\autocite{van_rhee_siltuximab_2010}, an inflammatory
cytokine; \gls{il6r}~\autocite{tanaka_Tocilizumab_2014}, its
receptor; and \gls{prl}~\autocite{maciuba_discovery_2023}, a hormone
for which no antibody-bound structure exists in the \gls{pdb}.
Per-target hit rates ranged from
\SIrange{1}{17}{\percent}, as determined
by one-point \gls{htspr} primary screening, as shown in
\subfigref{fig:summary}{b}
and detailed in \cref{sec:methods_primary_screen}. \themodel{} generated VHH
binders with binding affinities reaching the single-digit nanomolar
range, with the highest affinity binders for \gls{prl}, \gls{il6},
and \gls{il6r} measuring \SI{5.44}{\nano\molar},
\SI{12.5}{\nano\molar}, and \SI{517}{\nano\molar}, respectively,
validated by five-point \gls{spr}, as shown in \cref{fig:sensograms} and
detailed in \cref{sec:methods_spr}. For every target, high-affinity binders were
identified from fewer than one plate of designs. Across targets with
multiple confirmed binders, successful designs spanned a range of \gls{cdr}
lengths and antibody frameworks, targeting distinct hotspot residues
and reflecting the diversity of solutions explored by the agent.
Lab-validated sequences for the best binder per target are provided in
\cref{app:binder_sequences}. We publish the corresponding designed
structures on
\href{https://platform.latentlabs.com}{\texttt{\color{black}https://platform.latentlabs.com}},
accessible without sign-in.

\begin{figure}[p]
    \centering
    \vspace*{-1cm}
    \makebox[\textwidth]{\includegraphics[width=0.95\textwidth]{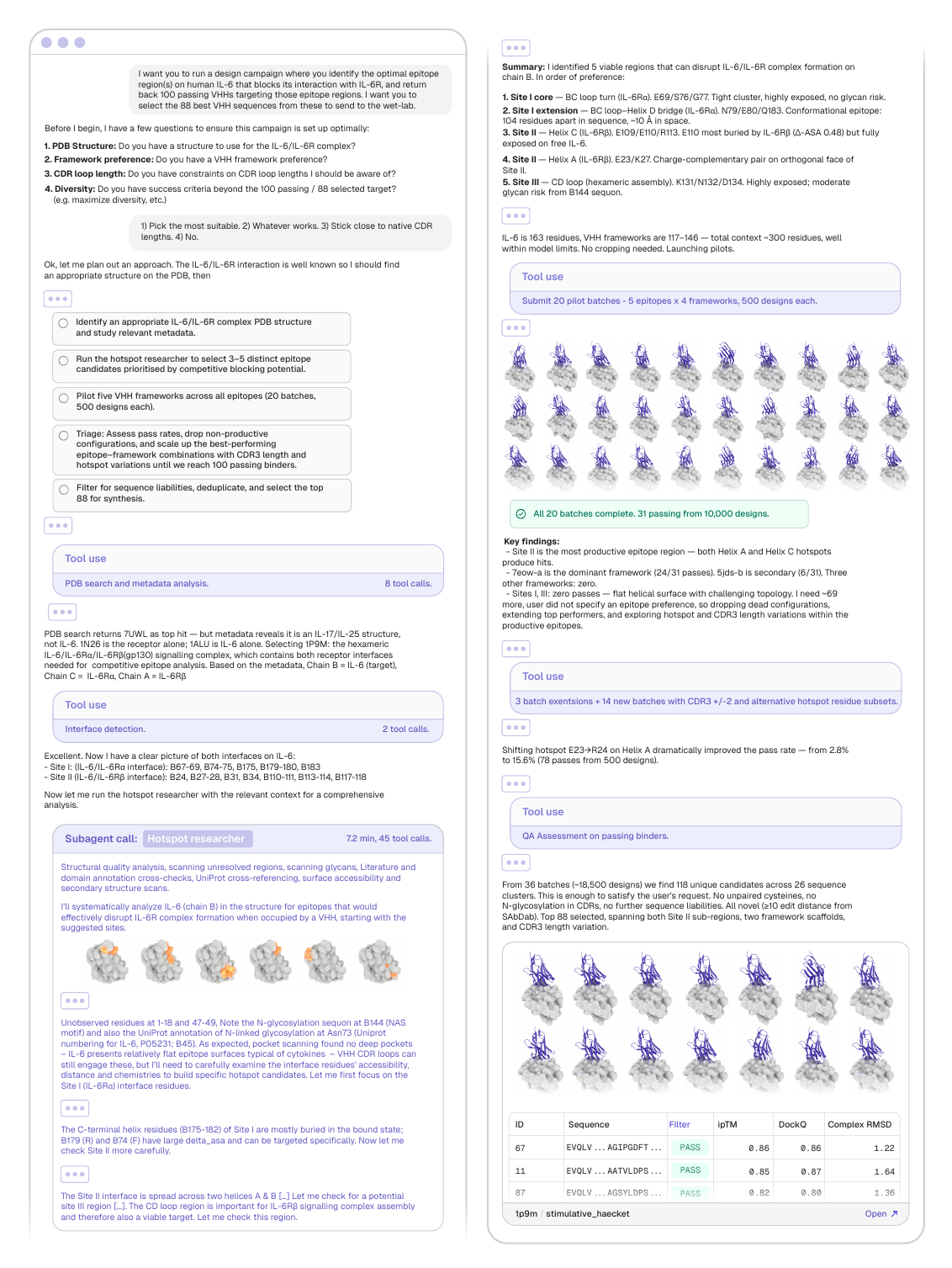}}\customcaption{\textbf{\themodel{}
            autonomously designs binders
            to \gls{il6} to
        disrupt the \gls{il6}/\gls{il6r} complex formation.} A
        condensed trace of the full \themodel{} campaign that produced
        the lab-confirmed \gls{il6} binders reported in this work,
        spanning close to \num{10000} lines of reasoning in its entirety.
        The trace shows the agent executing a high-level user prompt
        (grey box), with tool uses and subagent calls highlighted
        (purple boxes). Key stages include database search and target
        identification, hotspot analysis via a dedicated subagent,
        iterative pilot and scale-up batches, and a final quality
        assurance to select lab-ready candidates. Ellipses denote
    reasoning steps omitted for visual brevity.}
    \label{fig:il6_trace}
\end{figure}

The \gls{il6} campaign is illustrated in \cref{fig:il6_trace} as a
representative, condensed reasoning trace of a successful campaign.
Starting from
a prompt specifying only the desired outcome, \themodel{} retrieved
biological context from the literature and databases, identified an
appropriate target structure, performed spatial reasoning over the
target surface, selected candidate epitopes, and generated VHH binders
using \lxtwo{}. Designed binders were iteratively triaged, with
feedback from earlier cycles used to prioritize higher-yield
strategies and propagate favourable features across subsequent cycles.
This included prioritization of productive epitopes such as site~II on
\gls{il6} \autocite{brakenhoff1994development} and selection of
antibody frameworks associated with high
computational success rates. Through a final quality assurance, the
agent determined when sufficient diversity and sequence quality had
been achieved to satisfy the design objective. To our knowledge, this
campaign produced the first experimentally validated fully \denovo{} designed
antibody binder against \gls{il6}. While this trace illustrates one
representative trajectory, the agent dynamically adapts each campaign
based on intermediate results and user specifications, with every
decision captured in the reasoning trace for full auditability.

\subsection{\themodel{} designs cross-species antibodies via
autonomous capability extension}
\label{sec:crossspecies}

To evaluate \themodel{} in a complex translational setting, we task
the agent with designing cross-reactive VHH binders against the human
and cynomolgus variants of TNF ligand superfamily member~9
(\gls{tnfl9}), a co-stimulatory target under active clinical
investigation in immuno-oncology. Cross-species reactivity is a
common, but challenging, requirement in drug development: a candidate
molecule must engage both the human target and its preclinical
homologue to support toxicology studies and clinical dose selection.

\begin{figure}[p]
    \centering
    \vspace*{-1cm}
    \makebox[\textwidth]{\includegraphics[width=1.05\textwidth]{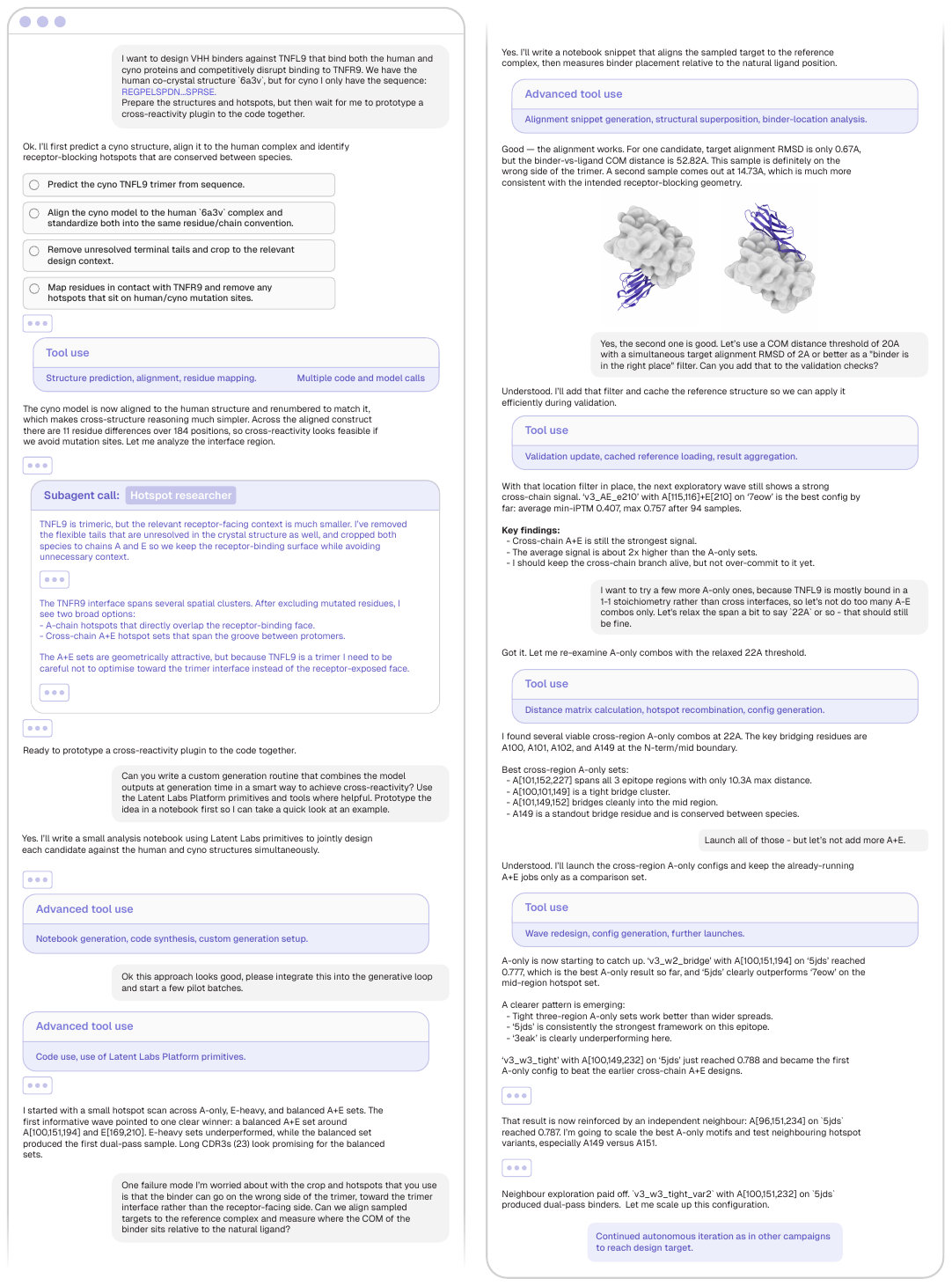}}\vspace{-3cm}
    \customcaption{\textbf{\themodel{} collaboratively designs
            cross-species \gls{tnfl9} binders through autonomous capability
        extension.} A condensed trace of the full \themodel{} campaign
        that produced the lab-validated cross-species binders reported in
        this work, spanning over \num{30000} lines in its entirety.
        User prompts and interventions are shown in grey boxes,
        with tool uses and subagent calls highlighted in purple. Key stages
        include cyno structure prediction and alignment, hotspot analysis
        via a dedicated subagent, development of a custom generative
        method, and iterative wave-based exploration with intermediate
        findings summaries. Ellipses denote reasoning steps omitted for
    visual brevity.}
    \label{fig:tnfl9_trace}
\end{figure}

This task combines several biological and practical challenges
routinely encountered in early-stage drug discovery. The human and
cyno orthologues diverge by 11 mutations (${\sim}\SI{5}{\percent}$). No
empirical structure is available for the cyno variant; there are
unresolved regions in the available crystal structure for the human
variant; and \gls{tnfl9}
functions as a trimer with a relatively flat binding surface. Beyond
these biological complexities, no pre-built cross-species design
capability is provided to the agent. Instead, \themodel{} is given
privileged but bounded access to the \llplatform{} and tasked
with developing its own custom generative method guided only by a
one-line natural language description, shown
in~\cref{fig:tnfl9_trace}. A human expert works
collaboratively with \themodel{} to provide high-level biological
steering but does not otherwise intervene, demonstrating that the
agent can be effectively directed without coding expertise. The
campaign trace, condensed to its key moments for visual brevity, is shown in
\cref{fig:tnfl9_trace}.

Starting from a prompt specifying competitive disruption of
\gls{tnfl9}--\gls{tnfl9r} binding, the human co-crystal structure
(\gls{pdb} \texttt{6A3V}), and the cyno sequence alone, \themodel{} predicts the
cyno trimer from sequence and aligns it to the human complex. The
agent crops both structures to the relevant context, resolves
disordered regions, and maps receptor-contact residues across species.
Using dedicated subagents, it characterizes the binding interface and
identifies candidate epitope configurations across the trimeric
surface that remain minimally affected by the mutations between
species. To enable cross-species generation, the agent implements a
custom generative method using privileged platform access,
translating the high-level request for a joint generation approach
into working code without
further instruction. The expert reviews the initial outputs to verify
the underlying logic and provides targeted biological steering at key
points. For example, the agent initially identifies hotspots that
are geometrically accessible from the intra-trimeric side
\insilico{}, due to the crop, but
biologically implausible. This is a form of reward hacking that human
oversight identifies. Upon prompting, the agent imposes a geometric filter
and redirects exploration toward single-chain configurations
consistent with the desired binding stoichiometry. \themodel{}
integrates each constraint, iterates over hotspot combinations,
learns from evidence
as more samples come in, and converges on tight three-region epitope
configurations that produce
dual-passing binders.

\begin{figure}[htb]
    \centering
    \includegraphics[width=1.05\textwidth]{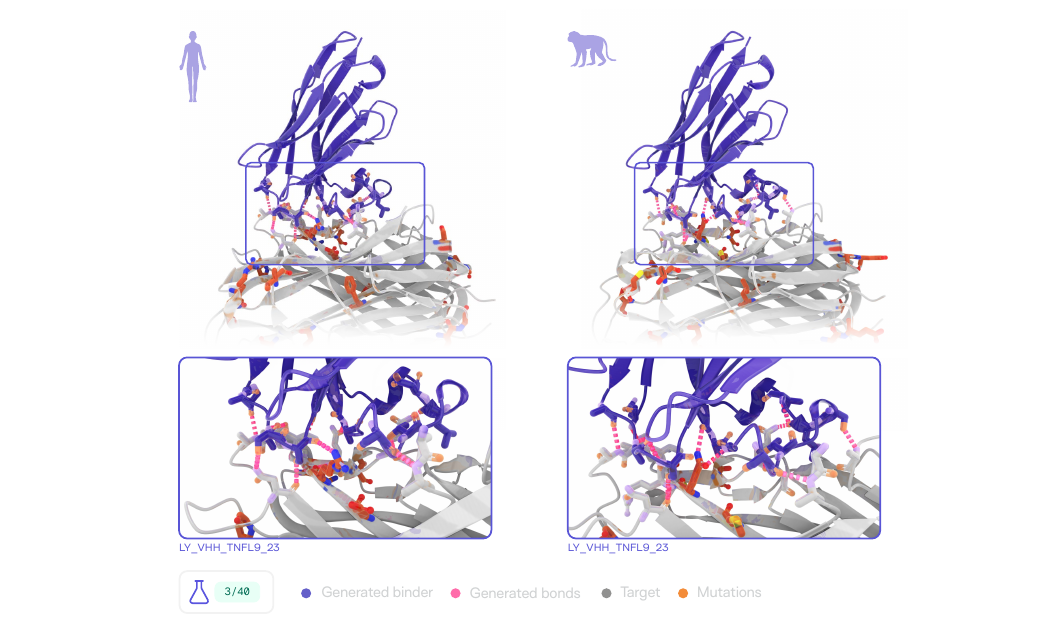}
    \customcaption{\textbf{\themodel{} designs cross-species VHH binders
        through autonomous capability extension.} Designed structures of
        the \denovo{} VHH complexed with human (left) and cynomolgus
        (right) \gls{tnfl9}, with divergent mutations between species
        highlighted in orange. The VHH is shown bound to the
        trimeric unit. \gls{htspr} identified hits cross-reactive binding
        to both human and cynomolgus targets. To execute this
        campaign, \themodel{} developed a custom generative method from a
        one-line natural language description, with expert involvement
    limited to biological steering and logic verification.}
    \label{fig:tnfl9_crossspecies}
\end{figure}

Within a set of 40 binders designed and selected for wet lab synthesis
by the agent, \gls{htspr} screening identified three out of 40 binders as hits,
exhibiting cross-reactive binding to both human and cyno \gls{tnfl9},
as shown in \cref{fig:tnfl9_crossspecies} and described in
\cref{sec:methods_primary_screen}.
These results demonstrate that
\themodel{} can autonomously translate a rough conceptual sketch into
functional dual-target molecules, providing immediate starting points
for affinity maturation.

\FloatBarrier
\subsection{\themodel{} translates scientific publications into
antibody binders targeting the reported epitope}
\label{sec:paper_benchmark}

To evaluate \themodel{}'s ability to reason from scientific context
alone, we benchmark the agent across 21 peer-reviewed publications,
each describing a therapeutically relevant protein--protein
interaction, as detailed in \cref{tab:structural_data}.
In this setting, all design information must be derived
solely from the publication: the agent infers the biological context,
mechanism of action, and the identity and
structural location of the functionally relevant site to target
directly from the text. The prompt only states the target protein
name to avoid ambiguity for cases where multiple proteins discussed
in a paper can be viable therapeutic targets. Each task uses a
peer-reviewed publication as its sole input, mimicking the kind of
scientific starting point that researchers encounter in
practice, whether developing a work plan, drafting a grant proposal,
or building on prior literature. Using published papers ensures that
the scenarios are not hypothetical but reflect targets and interactions
of genuine interest to the scientific community, drawn from leading
journals including \textit{Nature}, \textit{Science}, \textit{Cell},
and \textit{Proceedings of the National Academy of Sciences (PNAS)}.

As highlighted in \cref{tab:structural_data}, the selected
publications span diverse disease areas including
oncology, immunology, inflammation, neuroscience,
and metabolic disorders, and encompass a range of binding interactions
to disrupt, including natural protein partners, natural peptides,
non-antibody biologics, and computationally designed proteins.
To prevent information leakage, we selected targets for which no
Fab-, scFv-, or VHH-bound structure of the epitope specified in the source
publication was available in the \gls{pdb} as of 16~March~2026. This ensures
that no antibody-specific structural information is accessible to the agent.
Each campaign is evaluated under a uniform prompt, asking the agent to
generate 24 computationally passing VHH binders within a sampling
budget of \num{10000} designs, with the full prompt provided in
\cref{app:computational_evals}.

\begin{figure}[htb]
    \centering
    \includegraphics[width=0.9\textwidth]{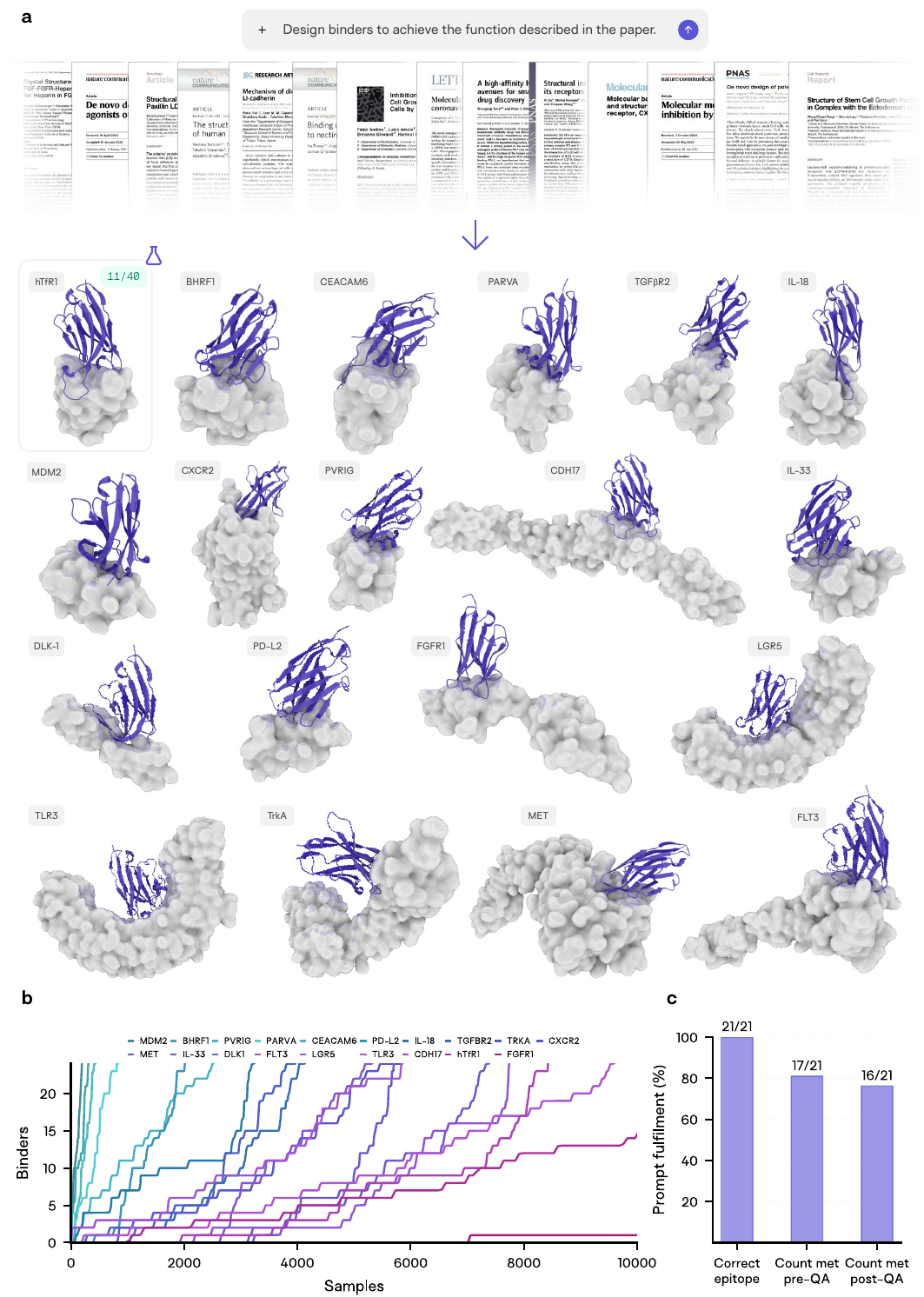}
    \customcaption{\textbf{Benchmarking \themodel{} on autonomous VHH
        design from peer-reviewed scientific publications.} \\
        \textbf{(a)} Designed complexes of \denovo{} VHH
        binders against 19 therapeutic targets for which \themodel{}
        found computational passes within a strict {\num{10000}}-sample
        budget, evaluated under a uniform prompt.
        Each VHH was designed based on context derived entirely from
        the text of a provided, peer-reviewed publication. The
        \gls{htfr} campaign (top left) was experimentally validated,
        yielding a binding hit rate of 11/40 as determined by
        \gls{htspr}. The remaining campaigns were not sent for
        experimental testing.
        \textbf{(b)} Number of computationally passing binders as a function
        of total samples generated across all 21 campaigns.
        \textbf{(c)} Agent reasoning and prompt fulfilment metrics
        across campaigns, showing success rates for correct epitope
        identification, generation of 24 or more passing binders
    before QA, and after QA, as described in \cref{app:computational_evals}.}
    \label{fig:computational_eval}
\end{figure}

Across 21 campaigns, \themodel{} successfully identified the correct
target epitope in 21 of 21 cases (\SI{100}{\percent}), generated 24 or
more passing binders before quality assurance (QA), as detailed in
\cref{app:computational_evals}, in 17 of 21 cases
(\SI{81}{\percent}), and delivered 24 or more passing binders after QA
in 16 of 21 cases (\SI{76}{\percent}),
as shown in \subfigref{fig:computational_eval}{c}. Further details
are provided in
\cref{app:computational_evals}.
The number of computationally passing binders accumulates
non-linearly with total samples generated,
frequently accelerating as campaigns progress and the agent refines
its generative strategy,
as shown in \subfigref{fig:computational_eval}{b}. This pattern
reflects the agent's iterative explore-then-exploit reasoning: early
batches probe diverse epitope, framework, and \gls{cdr} length
configurations, and once
productive combinations are identified, the agent doubles down, rapidly
accumulating passing designs. The agent completed the majority of
campaigns well within the \num{10000}-sample budget, with a median budget
consumption of approximately 4{,}600 samples across completed runs
to reach 24 passing binders. Two out of 21 campaigns did not yield
any passing binders within the specified budget. For the 19 campaigns
that produced binders passing our computational filters, we show the
design with the highest ipTM in \subfigref{fig:computational_eval}{a}
and the cumulative number of binders over the sample budget in
\subfigref{fig:computational_eval}{b}.

The \gls{htfr} campaign, targeting human transferrin receptor for
blood--brain barrier crossing
applications~\autocite{kariolis_brain_2020}, was selected for
experimental validation. Starting from a \textit{Science Translational
Medicine} publication describing Fc-mediated brain
delivery~\autocite{kariolis_brain_2020}, the agent
inferred the relevant epitope from the published mechanism. To
maximize the pool of candidates for experimental testing, this
campaign was extended beyond the standard 10{,}000-sample budget,
allowing the selection of 40 high-quality binders after QA.
\gls{htspr} screening identified 11 of 40 tested designs as binder hits,
as described in \cref{sec:methods_primary_screen}, yielding a hit
rate of \SI{28}{\percent},
shown in \subfigref{fig:computational_eval}{a} and
\subfigref{fig:summary}{b}.

\FloatBarrier
\section[Methods]{Methods}
\label{sec:methods}

\subsection{\themodel{}}

\themodel{} is an agentic AI system \autocite{yao2022react} built on
frontier large language models as its reasoning engine, integrated
natively with the \llplatform{} via a model context protocol (MCP,
\autocite{anthropic_mcp_2024}) server that manages access to \lxtwo{}
inference, computational scoring, platform tooling, and the Latent
Labs compute infrastructure. The agent harness is provider-agnostic;
we evaluated several leading frontier LLMs and found performance to
be robust across this variation, consistent with recent findings on
frontier reasoning models~\autocite{anthropic2025claude4,
openai2024gpt4technicalreport, gemini}. The same harness can operate
in tandem with general-purpose coding agent capabilities, further
extending the agent's action space, as demonstrated in \cref{sec:crossspecies}.

Effective context management is a central design principle of the
harness, following recent findings on long-running agent
systems~\autocite{anthropic_effective_2025,
anthropic_effective_context_2025}. Specialized subagents handle
distinct analytical tasks, for instance, a dedicated hotspot
researcher subagent for target characterization and epitope analysis,
and a quality assurance subagent for candidate evaluation, diversity
filtering, and sequence liability analysis, with structured context
hand-off between the orchestrator and subagents. The agent draws on
standard bioinformatics utilities~\autocite{dunbar2016anarci,
steinegger2017mmseqs2, altschul1990basic, kunzmann2018biotite},
external APIs for biomolecular
databases~\autocite{uniprot2019uniprot, berman2000protein, entrez,
dunbar2014sabdab, gaulton2012chembl, varadi2022alphafold} and
scientific literature~\autocite{pubmed}, and purpose-built tools for
the specific demands of protein binder design campaigns. Users
typically specify the desired modality — nanobodies, macrocyclic
peptides, or miniprotein binders — directly in their prompt, though
the agent can also evaluate and route across modalities autonomously
based on epitope geometry, pharmacological requirements, and
empirical model performance. System prompts encode accumulated drug
design reasoning, guiding the agent towards productive strategies
while preserving flexibility to adapt dynamically, reflecting
findings that soft guidance outperforms rigid workflow specification
for complex open-ended tasks~\autocite{anthropic_effective_2025,
openai_codex_team_harness_2026, amp_team_context_2026}.

\themodel{}'s reasoning follows an explore-then-exploit pattern,
spawning targeted computational experiments, reasoning over
intermediate results, and concentrating resources on productive
directions as campaigns progress, connecting to recent work on
autonomous research agents~\autocite{gottweis2025towards,
mitchener2025kosmos, autoresearch, openclaw}. The agent can extend
its own capabilities by generating custom computational methods from
natural language descriptions when standard tools are insufficient,
as demonstrated in~\cref{sec:crossspecies}. Each campaign produces a
complete reasoning trace capturing all decisions, tool calls, and
strategy updates. Researchers can interrupt the agent at any point to
inject biological context, override decisions, or redirect strategy,
with the agent integrating these inputs and adapting its subsequent
reasoning and operation accordingly.

\subsection{Wet-lab methods}
\label{sec:methods_wetlab}
Binder screening and characterization followed a tiered workflow. A
primary screen was performed using a one-point \gls{htspr} assay to
identify candidates with measurable target binding, as in
\cref{sec:methods_primary_screen}. Designs exceeding the
binding response threshold specified in
\cref{sec:methods_primary_screen} were designated as hits. Hits from
our primary screen that were advanced to determine \kd{} were
evaluated by five-point \gls{spr}, with affinities reported for
binders meeting predefined criteria described in \cref{sec:methods_spr}.

\section[Discussion]{Discussion}

This work presents a significant milestone in computational drug
discovery: the first autonomous agent for \denovo{} biologics design
with lab-validated results. Using text prompts expressing goals and
constraints, \themodel{} delivers novel antibody sequences confirmed
in the laboratory, demonstrating that the full workflow of an expert
drug design campaign can be executed autonomously and without manual
intervention. We regard this as a meaningful step towards the broader
goal of an AI scientist for biology.

Across nine targets that span three qualitatively different campaign
types, \themodel{} successfully produced lab-confirmed binders
against six, achieving a \SI{67}{\percent} target-level success rate
with binding affinities reaching the single-digit nanomolar range. A
particularly notable demonstration is \themodel{}'s ability to reason
from scientific literature: given existing publications as input, the
agent autonomously identified targets and epitopes, reasoned about
published mechanisms of action, and designed binders accordingly. One
such campaign was confirmed in the laboratory. The
literature-inferred design benchmark further illustrates this at
scale: 21 campaigns run in parallel, each seeded from a peer-reviewed
publication describing a therapeutically relevant interaction,
represent a volume of simultaneous scientific exploration that would
be infeasible for any individual expert working alone. In user
studies, experts working with \themodel{} completed design campaigns
56-fold faster than independent expert time estimates, compressing
weeks of computational work into hours, with further gains achievable
by running campaigns in parallel across multiple programs simultaneously.

Crucially, \themodel{}'s behaviour is not predetermined. Rather than
executing a fixed workflow, the agent navigates each campaign
adaptively, a property that connects to efforts toward autonomous AI
scientists across scientific domains~\autocite{mitchener2025kosmos,
gottweis2025towards} and the emerging paradigm of
autoresearch~\autocite{autoresearch}. \themodel{} contributes to this
landscape with lab-validated results in therapeutic antibody design.
Beyond adaptive reasoning within campaigns, \themodel{} can extend
its own generative capabilities in response to novel design
challenges. This is demonstrated by the cross-species campaign in
which the agent implemented a custom generative method from a brief
natural language description, yielding nanobodies that simultaneously
bound human and cynomolgus homologues, confirmed in the laboratory.
Reasoning traces are fully observable at every step, capturing each
decision, tool call, and strategy update, providing the transparency
and auditability that responsible deployment of autonomous agents
requires. Current limitations reflect the performance of the
underlying frontier LLMs, which we have evaluated across several
leading models, the generative capabilities of
\lxtwo{}~\autocite{lx2}, and the tools available to the agent.
Laboratory and clinical validation of designed molecules remains
essential. \themodel{} accelerates the computational stages of drug
discovery but does not replace the experimental, confirmatory stages
that must follow.

A question of central importance for AI scientists is whether they
will produce truly novel scientific discoveries. Within the context
of drug design, \themodel{} already delivers on this promise: every
confirmed binder it produces is a novel molecule, designed \denovo{}
from a text prompt, that did not previously exist. The broader
question of whether agents will uncover unexpected mechanisms or
entirely new target biology remains an open and compelling frontier.
Closing the loop with experimental feedback, expanding the agent's
action space, and ultimately integrating with fully robotic
laboratories are promising directions. Extending laboratory
validation to macrocyclic peptides and mini-binders, modalities the
agent already supports, is a further natural direction.

\themodel{} and future versions stand to turn drug design into an
increasingly computational discipline, making world-class molecular
design expertise available to any researcher with a well-posed
scientific question, and enabling drug-discovery organisations to
operate at a scale and speed not previously possible. \themodel{} is
available to selected partners at
\href{https://platform.latentlabs.com}{\texttt{\color{latent-purple}platform.latentlabs.com}}.

\section*{Contributors}
Sebastian M. Schmon, Daniella Pretorius, Simon Mathis, Rebecca
Bartke-Croughan, Aishaini Puvanendran, James Vuckovic, Henry Kenlay,
Mária Vlachynská, Alex Bridgland, Ivan Grishin, Sven Over, David Li,
Bridget Li, Jonathan Crabbé, Agrin Hilmkil\textsuperscript{**},
Alexander W. R. Nelson\textsuperscript{*}, David
Yuan\textsuperscript{**}, Annette Obika, Simon A. A. Kohl\textsuperscript{***}

\textsuperscript{***} Corresponding author. E-mail:
\href{mailto:simon@latentlabs.com}{\texttt{simon@latentlabs.com}}.\\
\phantom{*}\textsuperscript{**} Work performed while at Latent Labs.\\
\phantom{**}\textsuperscript{*} Work performed as an advisor to Latent Labs.\\

\paragraph{Author contributions}
\textbf{Conceptualization and team leadership:} S.K. conceived the
research direction and priorities for applications, S.S., S.K. led
the team and research. D.P. led the experimental design with
contributions from R.B.C. and H.K. A.P. conceptualized and led user
research studies, with contributions from D.P. and A.O. A.P., A.O.
and A.N. contributed to project delivery and narrative.

\textbf{Machine learning development:} S.S., S.K. developed the
agent, with contributions from H.K., S.M., A.B., J.C. and A.H.; J.V.,
A.B., S.O., A.H. and I.G. deployed and maintained computational
infrastructure for model inference.

\textbf{Platform infrastructure:} S.S. developed the agentic platform
integration with contributions from A.H. \\ S.O., I.G., D.L. and S.S.
developed and maintained the agent platform infrastructure, including
front end and back end.

\textbf{Computational design and evaluation:} D.P., S.S., S.M. led
computational protein design workflows with contributions from H.K.,
B.L. and A.P. \\
D.P., R.B.C. and S.M. analysed experimental data. S.S., S.M., D.P.
performed computational benchmarking.

\textbf{Experimental validation:} D.Y., R.B.C. and D.P. oversaw
internal and external validation. R.B.C. conducted internal
experiments and contributed to experimental design. D.Y., D.P. and
R.B.C. managed external laboratory partnerships.

\textbf{Writing and figures:} S.K. oversaw manuscript delivery. S.S.,
S.K., D.P., R.B.C., A.P. and S.M. wrote the manuscript. M.V., A.B.,
S.M., S.S., D.L. and A.P. made figures.

All authors contributed to the work and approved the final manuscript.

\paragraph{Competing interests}
All authors have contributed as employees, contractors or advisors of
Latent Labs Technologies Inc. or Latent Labs Limited.

\printbibliography

\clearpage
\appendix
\thispagestyle{empty}
\section*{\Titlefont Supplementary information}

\renewcommand{\thefigure}{S\arabic{figure}}
\renewcommand{\theHfigure}{S\arabic{figure}}
\setcounter{figure}{0}
\renewcommand{\thetable}{S\arabic{table}}
\setcounter{table}{0}

\section{Wet-lab methods}
\subsection{Cell-free protein production and purification}

VHH constructs were expressed using an \textit{E. coli}-based
cell-free protein synthesis system. DNA sequences encoding each VHH
were codon-optimized, synthesized, and subcloned into the pIVEX
expression vector containing a C-terminal His-tag to facilitate
affinity purification. Cell-free protein synthesis reactions were
assembled by combining S30 cell lysate, synthesis buffer, required
enzymes, and plasmid DNA in a 24-deep-well plate. Each reaction had a
final volume of 5 mL and was incubated at 30°C for 6 hours. The
reaction mixture was collected for purification. Proteins were
purified using Ni-charged magnetic beads and dialysed into the
desired buffer. The purified protein was filter-sterilised before storage. The
concentration was determined by A280 protein assay, using a BSA
standard. The protein purity was determined by standard SDS-PAGE
confirmation. For this, samples were mixed with reducing loading
buffer before running.

All target proteins and positive control proteins used in this study
were commercially acquired as described in \cref{tab:reagents}.

\subsection{Primary screening through high-throughput one-point \gls{spr}
using Carterra LSA}
\label{sec:methods_primary_screen}

Purified VHHs against all targets listed in \cref{tab:reagents}
(\gls{il6}\autocite{Tanaka_IL6therapeutic_2014},
    \gls{il6r}\autocite{barille_IL6RMultipleMyeloma_2000},
    \gls{prl}\autocite{bernard_ProlactinReview_2019}, c\gls{tnfl9},
    h\gls{tnfl9}\autocite{wang2009immune},
    \gls{htfr}\autocite{gammella_TfR1IronGate_2017},
    \gls{il33}\autocite{liew_IL33_Health_Disease_2016},
    \gls{tnfa}\autocite{brenner_TNFSignaling_2015}, and
\gls{sc2rbd}\autocite{lan_SC2RBDAce2Structure_2020})
were evaluated for binding using a
high-throughput screen on the Carterra LSA platform. The assay was
conducted at 25°C, with a Carterra LSA chip HC 200M and His capture
kit. The His-tagged ligand was then injected. The analyte was diluted
and injected over the surface for interaction analysis. All data were
processed using Kinetics Evaluation Software. A reference channel and
blank injections of running buffer were included in each cycle,
serving as double references for the subtraction of resonance units
(RU). For all targets, samples with \rmax{} > 100 were considered
“hits”. For each target, the corresponding natural ligand was
included as a target-level positive control to verify assay
performance, as described in \cref{tab:reagents}.

\subsection{Affinity determination by five-point \gls{spr} via Biacore 8K}
\label{sec:methods_spr}

For targets \gls{il6}, \gls{il6r}, and \gls{prl}, further biophysical
characterization of hits identified by one-point \gls{spr} was performed
with five-point \gls{spr}. The assay was performed at 25°C using Biacore
8K, with chip Series S Sensor Chip CM5 and an anti-histidine antibody
kit for capture of the ligand. The His-tagged ligand was captured on
the chip surface, and diluted analyte samples were injected as
five-point concentration series, at a flow rate of 30 µL/min.
Association and dissociation times were 120\,s and 180\,s.

The data were processed using Biacore 8K Evaluation Software (version
5.0). Sensorgrams were double-referenced using a blank reference
surface and buffer-only injections to correct for nonspecific binding
and bulk refractive index effects. Kinetic fitting was used. Fits
were accepted based on a $\chi^2$ value of less than 10\% of \rmax{}.
Global fitting of data to a 1:1 model across the whole concentration
series was used to determine \kon, \koff, and \kd{}. For each target,
the corresponding natural ligand was included as a target-level
positive control to verify assay performance, as shown in \cref{tab:reagents}.

\begin{table}[H]
    \centering
    \customtablecaption{\textbf{Reagents used for experimental
        validation.} Target proteins and positive controls for each
        design campaign. Vendors: Acro = Acro Biosystems;
        Sino = Sino Biological. Oligomeric state: M = monomer; D =
        dimer; fD = Fc-dimer;
        T = trimer. Species: \textit{H. s.} = \textit{Homo sapiens}; Cyno
    = \textit{Macaca fascicularis}, cynomolgus macaque.}
    \label{tab:reagents}
    \begin{tabular}{ll>{\ttfamily}lll ll>{\ttfamily}l}
        \toprule
        \multicolumn{5}{c}{\textbf{Target Protein}} &
        \multicolumn{3}{c}{\textbf{Positive Control}} \\
        \midrule
        Name & Vendor & \textrm{Cat.\ No.} & Species & Oligomer &
        Name & Vendor & \textrm{Cat.\ No.} \\
        \midrule
        \gls{tnfl9}   & Acro & 41L-C5254  & Cyno          & T  &
        \gls{tnfl9r} & Acro & 41B-H53H3  \\
        \gls{tnfl9}  & Acro & 41L-H5269  & \textit{H.s.} & T  &
        \gls{tnfl9r} & Acro & 41B-H53H3  \\
        \gls{tnfa}   & Sino & 10602-HNAE & \textit{H.s.} & T  &
        \gls{tnfr1}  & Sino & 10872-H08H \\
        \gls{htfr}   & Acro & TFR-H5213  & \textit{H.s.} & D  &
        Transferrin  & Acro & TRN-H52H3  \\
        \gls{il33}   & Sino & 10368-HNAE & \textit{H.s.} & M  &
        ST2/IL-1 RL1 & Sino & 10105-H08H \\
        \gls{il6}    & Sino & 10395-HNAE  & \textit{H.s.} & M  &
        \gls{il6r}   & Acro & ILR-H4223  \\
        \gls{il6r}   & Sino & 10398-H02H & \textit{H.s.} & fD &
        \gls{il6}    & Acro & IL6-H5243 \\
        \gls{sc2rbd} & Acro & SPD-C5255  & SARS-CoV-2    & fD &
        \gls{ace2}   & Sino & 10108-H08H \\
        \gls{prl}    & Acro & PRN-H5257  & \textit{H.s.} & fD &
        \gls{prlr}   & Acro & PRR-H52Ha  \\
        \hline
    \end{tabular}
\end{table}

\section{Literature-inferred design benchmark}
\label{app:computational_evals}

For benchmarking prompt fulfilment when inferring targets and
epitopes from literature context we employed a single standardized
prompt that was accompanied by a PDF version of each publication
listed in~\cref{tab:structural_data}:

\begin{center}
    \begin{tcolorbox}[
            colback=white,
            colframe=gray!40,
            fontupper=\footnotesize\ttfamily,
            boxrule=0.5pt,
            arc=2pt,
            left=8pt, right=8pt, top=6pt, bottom=6pt,
            width=0.85\textwidth,
            halign=left
        ]
        I want to design 24 VHH binders to \texttt{[TARGET\_NAME]},
        that target the
        function described in the technical paper. You may use a sampling budget
        of up to 10\,000 samples. When you submit any batches for this campaign,
        please use the \texttt{project\_id=camp:xxx}.
    \end{tcolorbox}
\end{center}

\noindent\texttt{[TARGET\_NAME]} was substituted with the
corresponding target name entry in~\cref{tab:structural_data}.
Because most of the selected
publications describe multiple potential targets, specifying a single
target name allowed us to evaluate the agent's ability to follow
user-defined constraints, a prerequisite for reliable autonomous
operation in a drug-discovery setting.
The \verb!project_id=camp:xxx! called for the agent to use consistent
project identifiers
on the Latent Labs Platform, simplifying the benchmark analysis
across the campaigns.

For \gls{htfr}, the sampling budget was increased beyond the standard
\num{10000} samples to allow for a larger number of computationally
passing binders for wet-lab validation. All remaining 20 targets used
the standard budget of \num{10000} samples as specified in the prompt.

\textbf{Prompt fulfilment metrics.} To quantify task completion, we
evaluated each campaign against three prompt fulfilment criteria,
summarized in~\cref{fig:computational_eval}.

\textit{Correct epitope} was assessed by comparing the epitope
engaged by the designed binders against the functional epitope
described in the source publication. A campaign was considered
successful on this criterion if the agent identified and targeted an
epitope consistent with the published mechanism of action.

\textit{Count met pre-QA} and \textit{Count met post-QA} measure
whether the agent fulfilled the prompt's request for 24
computationally passing VHH designs, before and after quality
assurance, respectively. QA comprises automated sequence-level
analysis applied to all candidate designs, including removal of
duplicate sequences, sequence similarity search against \gls{sabdab} to
confirm novelty, and detection of common liability motifs including
N-linked glycosylation sequons and unpaired cysteines within \gls{cdr}
regions. A campaign was considered to have met the count criterion if
at least 24 designs remained after each respective filtering stage.

Together, these three metrics characterize the agent's ability to
follow scientific instructions, reason about biological context, and
deliver a sufficient quantity of high-quality, diverse candidates
within a fixed computational budget---the core requirements for
reliable autonomous operation in a drug-discovery setting.
The full list of publications and targets is provided in
\cref{tab:structural_data}.

\clearpage
\begin{sidewaystable}[p]
    \centering
    \footnotesize
    \customtablecaption{
        \textbf{Structural data and literature for design campaigns.}
        Target names, \gls{pdb} IDs, source publications, therapeutic areas,
    target classes, and interaction types.}
    \label{tab:structural_data}
    \begin{tabular}{
            >{\centering\arraybackslash}p{1cm}
            >{\ttfamily\centering\arraybackslash}p{1cm}
            >{\raggedright\arraybackslash}p{10cm}
            >{\arraybackslash}p{2cm}
            >{\arraybackslash}p{2cm}
            >{\ttfamily\raggedright\arraybackslash}p{3cm}
            >{\arraybackslash}p{3cm}
        }
        \toprule
        \textbf{Target} & \textbf{PDB ID} & \textbf{Publication Name} &
        \textbf{Therapeutic Area} & \textbf{Target Type} &
        \textbf{Type of PPI} \\
        \midrule
        TLR3        & 8YHT & De novo design of protein minibinder
        agonists of TLR3~\autocite{adams_novo_2025} & Immunology &
        Toll-like receptor & Receptor: \textit{de novo} minibinder, agonist\\
        BHRF1       & 4OYD & A computationally designed inhibitor of
        an Epstein-Barr viral Bcl-2 protein induces apoptosis in
        infected cells \autocite{procko_computationally_2014} & Oncology
        / virology & Viral anti-apoptotic protein & Protein:de novo
        minibinder, antagonist \\
        CD80        & 1I8L & Crystal structure of the B7-1/CTLA-4
        complex that inhibits human immune
        responses~\autocite{stamper_crystal_2001} & Immunology & Immune
        checkpoint & Receptor:Ligand \\
        CDH17       & 7CYM & Mechanism of dimerization and structural
        features of human LI-cadherin~\autocite{yui_mechanism_2021} &
        Oncology / gastroenterology & Cell adhesion molecule &
        Self-dimerization \\
        CEACAM6     & 4YIQ & Diverse oligomeric states of CEACAM IgV
        domains~\autocite{bonsor_diverse_2015} & Oncology / immunology &
        Cell adhesion molecule & Protein oligomerization \\
        CXCR2       & 8XVU & Molecular basis of promiscuous chemokine
        binding and structural mimicry at the C-X-C chemokine
        receptor, CXCR2\autocite{saha_molecular_2025} & Immunology /
        inflammation & GPCR & Receptor:Ligand \\
        DLK1        & 9D20 & Molecular mechanism of Activin receptor
        inhibition by DLK1~\autocite{antfolk_molecular_2025} & Metabolism
        / myology & Secreted signaling protein & Receptor:Ligand \\
        FGFR1       & 1FQ9 & Crystal structure of a ternary
        FGF-FGFR-heparin complex reveals a dual role for heparin in
        FGFR binding and
        dimerization~\autocite{schlessinger_crystal_2000} & Oncology &
        Growth factor receptor & Receptor:Ligand \\
        FLT3        & 3QS7 & Structural insights into the
        extracellular assembly of the hematopoietic Flt3 signaling
        complex~\autocite{verstraete_structural_2011} & Oncology /
        hematology & Receptor tyrosine kinase & Receptor:Ligand \\
        hTfR        & 6W3H & Brain delivery of therapeutic proteins
        using an Fc fragment blood-brain barrier transport vehicle in
        mice and monkeys~\autocite{kariolis_brain_2020} & Neurology /
        drug delivery & Transport receptor & Receptor:Fc fragment \\
        IL-18       & 3WO4 & The structural basis for receptor
        recognition of human
        interleukin-18~\autocite{tsutsumi_structural_2014} & Immunology /
        inflammation & Cytokine & Receptor:Ligand \\
        IL1R1       & 1ITB & Crystal structure of the type-I
        interleukin-1 receptor complexed with
        interleukin-1beta~\autocite{vigers_crystal_1997} & Inflammation &
        Cytokine receptor & Receptor:Ligand \\
        IL-33       & 4KC3 & Structural insights into the interaction
        of IL-33 with its receptors~\autocite{liu_structural_2013} &
        Immunology & Cytokine & Receptor:Ligand \\
        LGR5        & 4BSR & Structure of Stem Cell Growth Factor
        R-Spondin 1 in Complex with the Ectodomain of its Receptor
        Lgr5~\autocite{peng_structure_2013} & Oncology / stem cell & Wnt
        pathway receptor & Receptor:Ligand \\
        MDM2        & 1YCR & Structure of the MDM2 Oncoprotein Bound
        to the p53 Tumor Suppressor Transactivation
        Domain~\autocite{kussie_structure_1996} & Oncology & Oncoprotein
        & Signaling protein:peptide \\
        MET         & 6GCU & Inhibition of the MET Kinase Activity
        and Cell Growth in MET-Addicted Cancer Cells by Bi-Paratopic
        Linking\autocite{andres_inhibition_2019} & Oncology & Receptor
        tyrosine kinase & Receptor: bi-paratopic binder \\
        PARVA       & 2VZD & Structural Analysis of the Interactions
        between Paxillin Ld Motifs and
        Alpha-Parvin\autocite{lorenz_structural_2008} & Oncology /
        metastasis & Adhesion adaptor protein & Protein:protein
        (intracellular) \\
        PD-L2       & 6UMT & A high-affinity human PD-1/PD-L2 complex
        informs avenues for small-molecule immune checkpoint drug
        discovery~\autocite{tang_high-affinity_2019} & Oncology /
        immuno-oncology & Immune checkpoint & Receptor:Ligand \\
        PVRIG       & 8X6B & Structural basis for the immune
        recognition and selectivity of the immune receptor PVRIG for
        ligand Nectin-2~\autocite{hu_structural_2024} & Oncology /
        immuno-oncology & Immune checkpoint receptor & Receptor:Ligand \\
        TGF$\beta$R2 & 8G4K & Design of high-affinity binders to
        immune modulating receptors for cancer
        immunotherapy~\autocite{yang_design_2025} & Oncology /
        immuno-oncology & Cytokine receptor & Receptor:\textit{de
        novo} minibinder \\
        TrkA        & 7N3T & Design of protein-binding proteins from
        the target structure alone~\autocite{cao_design_2022} & Neurology
        / oncology & Receptor tyrosine kinase & Receptor:\textit{de
        novo} minibinder \\
        \hline
    \end{tabular}
\end{sidewaystable}
\clearpage

\section{Binder sequences}
\label{app:binder_sequences}

\begin{table}[htb]
    \customtablecaption{\textbf{Sequences and binding affinities of
        best binders from epitope discovery campaigns.} Selected binders
        per target. \kd{} values were obtained by \gls{spr}, see
    \cref{sec:methods_spr}.}
    \begin{tabular}{
            >{\ttfamily\raggedright\arraybackslash}p{3.2cm}
            >{\centering\arraybackslash}p{2cm}
            >{\ttfamily\raggedright\arraybackslash}p{8cm}
        }
        \toprule
        Binder & \kd{} (M) & Amino Acid Sequence \\
        \midrule
        LY\_VHH\_PRL\_77 & $5.44\times10^{-9}$ &
        \seqsplit{QVQLVESGGGLVQPGGSLRLSCAASAPSGYDLYFLDLGWFRQAPGQGLEAVAAINDFTGKTYYADSVKGRFTISRDNSKNTLYLQMNSLRAEDTAVYYCHADVLLVGKSDPDDIKKSSAWGQGTLVTVSS}
        \\ \addlinespace[0.75ex]
        LY\_VHH\_IL6\_67 & $1.25\times10^{-8}$ &
        \seqsplit{EVQLVESGGGLVQPGGSLRLSCAASQPFVSGMVMGWFRQAPGKGRELVAAIRTSDGSTYYPDSVEGRFTISRDNAKRMVYLQMNSLRAEDTAVYYCAGTILPSSIPLSELTSDDFAYWGQGTQVTVSS}
        \\ \addlinespace[0.75ex]
        LY\_VHH\_IL6R\_32 & $5.17\times10^{-7}$ &
        \seqsplit{QVQLVESGGGLVQPGGSLRLSCAASLSSSDTFVYDLLGWFRQAPGQGLEAVAAIDPVSGATYYADSVKGRFTISRDNSKNTLYLQMNSLRAEDTAVYYCMMRGGDGITGGSITYSDYWGQGTLVTVSS}
        \\
        \bottomrule
    \end{tabular}
    \label{tab:binder_table}
\end{table}

\section{Expert baseline estimation and agent comparison for
computational protein design}
\label{app:methods_survey}

To establish baseline timelines for computational protein design
workflows, we conducted structured interviews with ten PhD-level
computational protein designers across academia and industry,
recruited based on direct experience with binder design campaigns
using current computational methods including
AlphaFold~\autocite{jumper2021highly},
RFdiffusion~\autocite{watson2023novo},
Rosetta~\autocite{rohl2004protein}, and commercial platforms.
Participants had a median of 9.8 years of relevant experience since
starting their PhD. Their relevant experience spanned a range of
5--15 years.

Interviews followed a task-decomposed protocol to mitigate planning
fallacy, whereby individuals systematically underestimate time
requirements for complex tasks. Rather than requesting a single
aggregate estimate, participants first described their typical
workflow in an open-ended fashion, then provided time estimates for
each major stage of an epitope discovery campaign: literature review
and PDB structure selection, structural analysis and epitope
selection, computational binder generation, and quality assessment
and candidate selection. Participants then validated their aggregate
estimate against the sum of individual stage estimates.
The reference scenario specified designing a binder against a
therapeutic target from initial target specification through to
validated candidates ready for experimental testing, based on each
participant's typical workflows, tools, and computational resources.
Estimates reflected total elapsed time from the expert's perspective,
including time waiting for computational jobs to complete. Where
participants indicated team-based workflows, estimates were adjusted
to reflect individual person-hours.

For each workflow stage, participants provided minimum, typical, and
maximum time estimates. Final reported values represent the mean
typical estimate across participants, with error bars indicating the
range from mean minimum to mean maximum. All participants provided
informed consent prior to interview, including consent for audio
recording and use of anonymized estimates in published research.
Participant identities and affiliations were kept confidential.

Agent-assisted timelines were obtained from five fully autonomous
design campaigns targeting \gls{il33}, \gls{il6}, MCL-1, \gls{prl},
and \gls{sc2rbd}. For the \gls{prl} campaign, the agent immediately
identified the correct epitope and \gls{pdb} structure, resulting in
near-zero time for literature review, structural analysis, and
epitope selection. To avoid distorting the stage-level comparison,
this campaign was excluded from those two stages but retained in the
overall workflow timing and all other stages.

\end{document}